# Global maps of Venus nightside mean infrared thermal emissions obtained by VIRTIS on Venus Express

A. Cardesín-Moinelo (1,2)*, G. Piccioni (3), A. Migliorini (3), D. Grassi (3), V. Cottini (4,5), D. Titov (6), R. Politi (3), F. Nuccilli (3), P. Drossart (7)

(1) European Space Astronomy Centre, Villanueva de la Cañada, Madrid, Spain
(2) Instituto de Astrofísica e Ciências do Espaço, Obs. Astronomico de Lisboa, Portugal
(3) Istituto di Astrofisica e Planetologia Spaziali, Istituto Nazionale Astrofisica, Rome, Italy
(4) University of Maryland, College Park, Maryland, USA
(5) NASA Goddard Space Flight Center, Greenbelt, Maryland, USA
(6) European Space Research and Technology Center, Noordwijk, The Netherlands
(7) LESIA, Observatoire de Paris, Meudon, Paris, France

* All correspondence to: Alejandro.Cardesin@esa.int

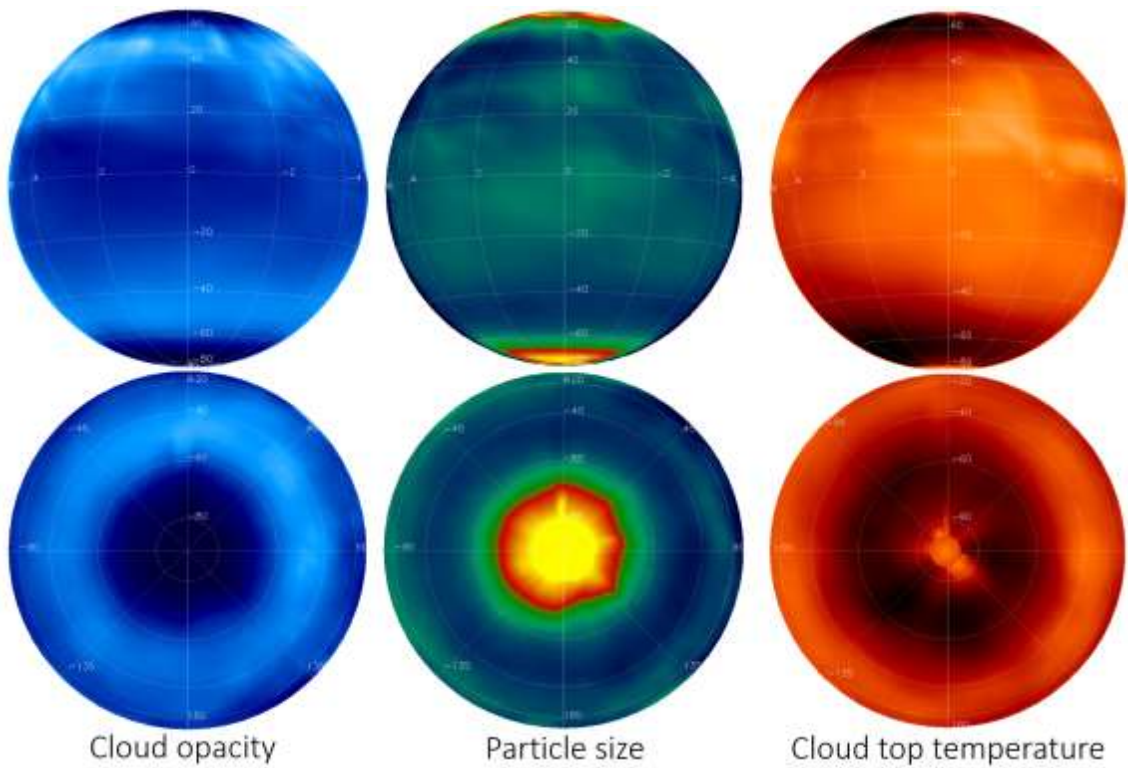

**ABSTRACT**

One of the striking features about Venus atmosphere is its temporal variability and dynamics, with a chaotic polar vortex, large-scale atmospheric waves, sheared features, and variable winds that depend on local time and possibly orographic features. The aim of this research is to combine data accumulated over several years and obtain a global mean state of the atmosphere focusing in the global structure of the clouds using the cloud opacity and upper cloud temperatures.

We have first produced global maps using the integrated radiance through the infrared atmospheric windows centred around 1.74µm and 2.25µm, that show the spatial variations of the cloud opacity in the lower clouds around 44-48 km altitude and also provide an indirect estimation of the possible particle size. We have also produced similar global maps using the brightness temperatures seen in the thermal region at 3.8µm and 5.0µm, which provide direct indication of the temperatures at the top of the clouds around 60-70 km altitude.

These maps have been generated using the complete dataset of the Visible and InfraRed Thermal Imaging Spectrometer mapping channel (VIRTIS-M) on board Venus Express, with a wide spatial and long temporal coverage in the period from May 2006 until October 2008.

Our results provide a global view of the cloud opacity, particle size and upper cloud temperatures at both hemispheres, showing the main different dynamical regions of the planet. The profiles obtained also provide the detailed dependencies with latitude, local time and longitude, diagnostic of the global circulation flow and dynamics at various altitude layers, from about 44 up to 70 km over the surface.
- 2 -

# 1. INTRODUCTION

## 1.1. VENUS EXPRESS

Venus Express is the ESA's first mission to Earth's nearest planetary neighbour, launched on November 9th 2005 from Baikonur and arrived at its destination on April 11th 2006. The spacecraft was successfully inserted on a stable elliptical orbit of 24-hour period, with the pericenter at about 250 km of altitude near the north pole and the apocenter at about 66,000 km of altitude over the southern hemisphere. Being designed in this way, one spacecraft orbit is completed approximately in one Earth day, and hence the total number of orbits is indicative of the days from the orbit insertion of Venus Express. Among the main science goals of the mission, the study of the complex dynamics and structure of the atmosphere was one of the key objectives, as well as the interaction with the surface and outer space. [Svedhem 2007, Titov 2006]

## 1.2. VIRTIS INSTRUMENT

VIRTIS (Visible and Infra-Red Thermal Imaging Spectrometer) was the imaging spectrometer on-board Venus Express with the capability to map the planet from the surface up to the thermosphere. An imaging spectrometer of this type is perfectly suited to cover very different fields, from meteorology of the mesosphere to geology, through the near infrared deep windows sounding down to the surface [Drossart 2007]. VIRTIS was constituted of three channels in one compact instrument [Piccioni 2008]. One channel was devoted solely to spectroscopy at high resolution, housed in the High resolution (-H) optical subsystem (2-5µm). The other channels were housed in the Mapper (-M) optical subsystem and devoted to spectral mapping, with one detector in the visible (-M-VIS, 0.25-1.0µm) with a spectral sampling of about 2nm and another one in the infrared band (-M-IR, 1.0-5.2µm) with a spectral sampling of about 10nm. In this study we use only the VIRTIS-M-IR data, calibrated in spectral radiance units, with an uncertainty below $10^{-3}$ W/m$^2$/sr/µm in the spectral range between 1.7 µm and 5.1 µm [Piccioni 2008, Cardesin-Moinelo 2010].

## 1.3. VENUS LOWER CLOUDS IN THE INFRARED ATMOSPHERIC WINDOWS

Observations at different wavelengths, including ultra-violet (UV), visible (VIS) and near-infrared (NIR) spectral ranges, and comparison of observed features are very important to investigate cloud properties. As largely demonstrated in the past and more recently shown with the Akatsuki data [Limaye 2018a], observations in the NIR bands are able to probe clouds at lower altitudes with respect to the observations in the UV because these are less sensitive to scattering by small particles. However, the NIR $CO_2$ transparency windows can contribute to probe deeper in the atmosphere, and hence when same structures are found in the UV, VIS and NIR these might indicate that the features extend from the cloud top to a depth of one scale height and larger cloud particles are detected. The wavelengths analysed in the present paper, centred around 1.74 and 2.25µm, belong to radiance coming mainly from below the cloud layers. Radiative transfer models indicate that the radiance at these bands originates about 10–20 and 20–30 km above the surface, respectively, see [Allen and Crawford 1984, Crisp 1991, Carlson 1991] and also the recent review by [Titov 2018]. This thermal radiation from the lower layers goes up through the cloud layer and is attenuated mostly by the lower clouds with differing optical depths [Titov 2018, Sánchez-Lavega 2017]. Therefore, the features observed at these wavelengths show mainly the spatial variations in the opacity of the lower to middle clouds, around 44-48 km altitude. [Peralta 2017a]

These clouds have been studied since decades by Earth telescope observations, see [Tavenner 2008] and references therein and by several space missions, starting with the fly-by of Venus from the Galileo NIMS instrument [Carlson 1991], later by VIRTIS on Venus Express [Sánchez-Lavega 2008, Hueso 2012] and more recently by Akatsuki [Nakamura 2016] with the IR2 camera as described by [Satoh 2016, 2017] and [Peralta 2018], which also includes a good overview of Venus cloud observations at 2.3µm since 1978. These clouds are known to be composed mainly of sulphuric acid covering the whole planet [Marcq 2018, Titov 2018], dynamically elongated by the strong zonal winds coming from the super-rotation and travelling towards the poles with the meridional winds caused by the Hadley cell [Sánchez-Lavega 2018]. The general structure and dynamics of the clouds have been analysed and compared extensively with General Circulation Models [Lebonnois 2010, Garate-Lopez 2018, Kashimura 2019, Ando 2019].

Ground-based observations pointed out variability in the cloud distribution, the equator region being more cloudy than at mid-latitudes [Crisp 1991], and confirmed by more recent investigation that report a minimum of cloud opacity at mid-latitudes [Tavenner 2008]. The same conclusions are obtained with the Akatsuki IR2 camera at 2.3µm through investigation of the wind speed [Peralta 2018]. The authors report high opacity at low latitudes as well as complex cloud features at mid-latitudes. These dark and bright structures could be explained with horizontal inhomogeneities in the clouds' opacity, as already suggested by [Allen 1987] and [Crisp 1989].

The comparison of the different opacities observed at 1.74 µm and 2.3 µm has also been studied to provide an indication of the cloud particle size, showing a uniform spatial distribution in the low to middle latitudes, with a significant increase of the particle size towards the poles [Carlson 1993, Grinspoon 1993, Wilson 2008, McGouldrick 2008]. The comprehensive retrieval by [Haus 2013] also reported a minimum of the particle size around the mid-latitudes, concluding that mid latitude clouds between 30 and 60º are covered by thiner clouds than the equatorial and polar latitudes.

## 1.4. VENUS CLOUD TOP TEMPERATURES SEEN IN THE THERMAL REGION

In the second part of this study, we focus on the upper cloud layer that is observed by VIRTIS-M in the thermal range from 3 to 5µm, around the main $CO_2$ absorption band at 4.3µm, visible in Figure 5. This wavelength region has been discussed at length in the recent review by [Limaye 2018b] and studied by several groups in the past years, including various specific analysis using VIRTIS data. In particular [Grassi 2008, 2010, 2014] provided average temperature fields focusing mostly in the southern hemisphere and describing the local time variabilities and strong dusk/dawn asymmetries in the cloud top temperatures especially around the south cold collar region. [Garate-Lopez 2015] extended the analysis of this local time variability around the southern cold collar, including an extensive study of the strong variations in altitude, shape and very fine structures around the south polar vortex. [Migliorini 2012] provided temperature field variations using VIRTIS-H data and covering a wider latitude range, reaching the equatorial region and northern hemisphere, although limited by the low resolution at high northern latitudes. [Haus 2013] also performed an independent analysis of the cloud top temperatures with VIRTIS and showed a remarkable latitudinal symmetry of both hemispheres with respect to the equator.

In this work we focus on the thermal brightness measured outside the main $CO_2$ absorption band, at two specific wavelengths around 3.8µm and 5.0µm, shown in Figure 6, where the atmospheric opacity from the cloud tops toward space is relatively low. Radiative transfer calculations set the main source of the radiance at these wavelengths as coming from a layer at about 60-70 km altitude, corresponding to the upper cloud layer [Peralta 2017a, Limaye 2018b]. Moreover, these wavelengths are of particular interest because they allow the study of the polar vortex morphology and dynamics. [Piccioni 2007, Luz 2011, Garate-Lopez 2013, 2015 and 2016]

The paper is organized as follows. Data are described in section 2. Results for the lower clouds are presented in section 3, while cloud top temperature maps are presented in section 4. Discussion and conclusion follow in section 5 and 6.

## 2. DATA SELECTION

This study is based on all VIRTIS-M-IR data from the beginning of the Venus nominal science mission in May 2006, starting with the first science orbit number 23. Previous observations obtained in April 2006 during the Venus orbit insertion and commissioning phase have not been included in this study as the instrument temperatures were much higher than the nominal values, affecting the spectral and radiometric calibration which could introduce important bias in the mean values.

A total of 3133 nightside observations have been included in this study, up to October 2008 when the VIRTIS-M-IR operations terminated due to the failure of the cryocooler, which was used to keep the cold temperature required by the infrared sensor.

The radiance from Venus at different wavelengths is very different, so there is no unique exposure time covering the full spectral range in an optimized way. The environmental temperature conditions are quite severe and a relatively high thermal background from the instrument dark current prevents the use of long exposure time without saturation [Piccioni 2008, Cardesín-Moinelo 2010]. On the other hand, the radiance from the atmospheric windows is very weak and it requires relatively long exposure time. The VIRTIS operations used an optimized exposure time depending on the science case for each specific observation. The long exposures, typically more than 3.3s, have been used roughly in ~60% of the nightside observations to focus in the deep atmosphere (windows 1.74 and 2.25µm). With such an exposure time, the thermal band saturates and hence thermal investigation (including cloud top temperature and atmospheric structure) can be addressed only below 4.2µm. The short exposures, below 0.36s, have been used in ~40% of the nightside observations and allow measuring the thermal region at longer wavelengths up to 5.1µm. The maps are constructed using all available data for each wavelength, including both long and short exposure times where possible, only excluding saturated data, in particular for the thermal region at 5.0µm where only short exposures have been used.

All input data are obtained directly from the VIRTIS calibrated data set corresponding to the nominal mission phase and first mission extension: [VEX-V-VIRTIS-2-3-V3.0] and [VEX-V-VIRTIS-2-3-EXT1-V2.0]. Input calibrated data are given in spectral radiance units (W/m$^2$/sr/µm) as described by [Cardesín-Moinelo 2010].

## 2.1. VIRTIS M-IR COVERAGE

The full coverage of all VIRTIS M-IR observations between May 2006 and October 2008, corresponding to orbits from 23 to 921, is shown in Figure 1 and Figure 2, distributed along local time and longitude respectively. The coverage distribution provides the visual statistical meaning of the data population for the different location of Venus.

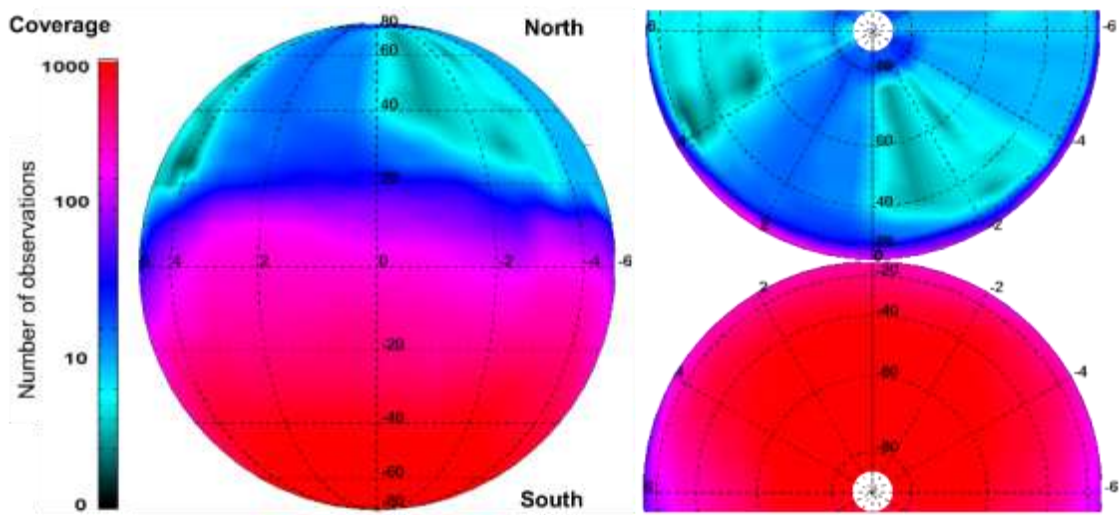

Figure 1. Local Time coverage of VIRTIS-M-IR observations (in Logarithmic scale) of Venus nightside, as seen from the equator at midnight (left) and from both poles (right).

Shown in an orthographic projection distributed along latitude (north up) and local time with a fixed grid of [5 degrees x 30 minutes].

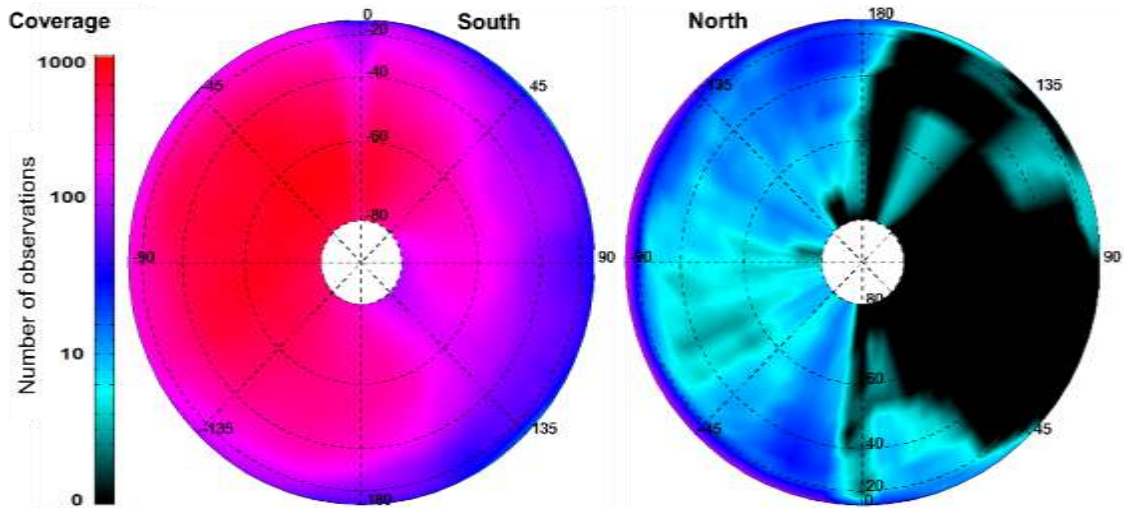

Figure 2. Longitudinal coverage of VIRTIS-M-IR observations (in Logarithmic scale) over both Venus hemispheres, as seen from the South pole (left) and North pole (right). Shown in an orthographic projection distributed along latitude and longitude (increasing eastward) with a fixed grid of [5 x 5 degrees].

As one can see, the southern hemisphere has very good coverage with several hundred observations per sample bin, whereas the coverage over the northern hemisphere is more patchy, with typically less than 10 observations per bin and many gaps especially in the longitudinal coverage. The dense coverage of the southern hemisphere is caused by the apoapsis of the Venus Express orbit, which reaches its highest altitude (~66,000km) near the south pole. In nominal observation modes, the instrument Field Of View (FOV) is typically a square of 64x64 mrad, which means that one single cube of observation can typically cover a projected square of ~4000x4000km as seen from the apoapsis. Indeed, at this distance the whole planet can be covered simply with nine observations, usually known as a "mosaic" and very common in the operations planning. [Titov 2006]

The more limited coverage in the northern hemisphere shows some of the gaps with regions that have never been observed during the mission. The pericenter situated over the North pole implies a spacecraft altitude very low over the northern hemisphere, spanning from 5000km at equator down to 250km or less near the north pole. At these low distances, the 64mrad wide FOV covers only 320km for a nadir observation at the equator and only 16km near the pericenter, with relatively high spatial resolution. In the northern hemisphere, the high relative speed of the spacecraft with respect to Venus, prevents the use of the scanning mirror to build an image, due to the low persistence of the target. For this reason, the observations near the pericenter are usually performed in push-broom mode, with the scanning mirror fixed at the boresight central position, pointing toward nadir or nearby, with increasing speed as it approaches the pericenter. This makes the footprints at the northern hemisphere appear like narrow stripes very extended in latitude.

Note also that there is a bias in the east-west longitudes which appear both in the northern and southern hemispheres, especially with big coverage gaps in the eastern-northern longitudes [0º-180ºE]. These gaps are even larger if we consider only the low exposure observations, which are roughly 40% of the total nightside observations and will impact the maps produced at 5.0µm as we will see later. Coverage gaps are constrained by the planetary and spacecraft ephemeris, all together impacting the observations during certain periods. Unfortunately, these longitude coverage gap could not be fulfilled with the Infrared channel observations.

## 2.2. DATA SELECTION FOR INFRARED ATMOSPHERIC WINDOWS

The radiance measured at 1.74µm is a direct measurement of cloud opacity [Grinspoon 1993, Titov 2018] and so the whole band can be used for this analysis. However, the window around 2.3µm contains important secondary components from the CO distribution and other chemicals ($H_2O$, OCS, $SO_2$) [Marcq 2008]. During our analysis we have performed various tests using different band widths to consider the effect of various species and we have seen that the impact of these trace gases is negligible for our study. In any case, to avoid any potential ambiguity, we focus our results only in the region 2.18-2.30µm that is dominated by $CO_2$ as proposed by [Tsang 2008 and 2009]. Note also the advantage that the radiance integrated over the two proposed windows has comparable values and so they can be shown using the same scale as in Figure 4.

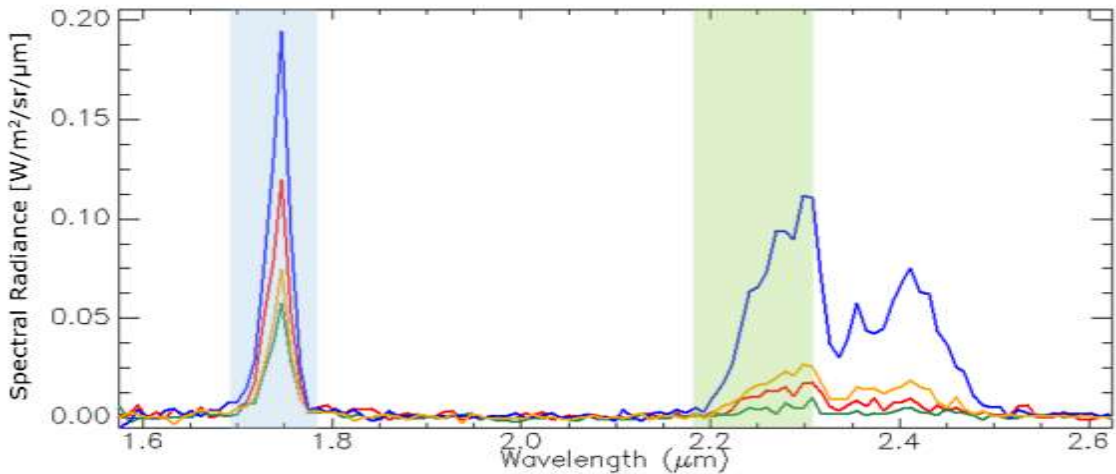

Figure 3. Typical VIRTIS-M infrared spectrum in the nightside of Venus between 1.6 and 2.6µm showing the main two atmospheric windows centred at 1.74µm (marked in blue) and 2.25µm (marked in green). The different coloured plots show typical variations in the radiances seen through these windows at various locations of the Venus nightside atmosphere: Red = polar vortex; Green = cold collar, Blue = mid-latitude, Orange = low-latitude

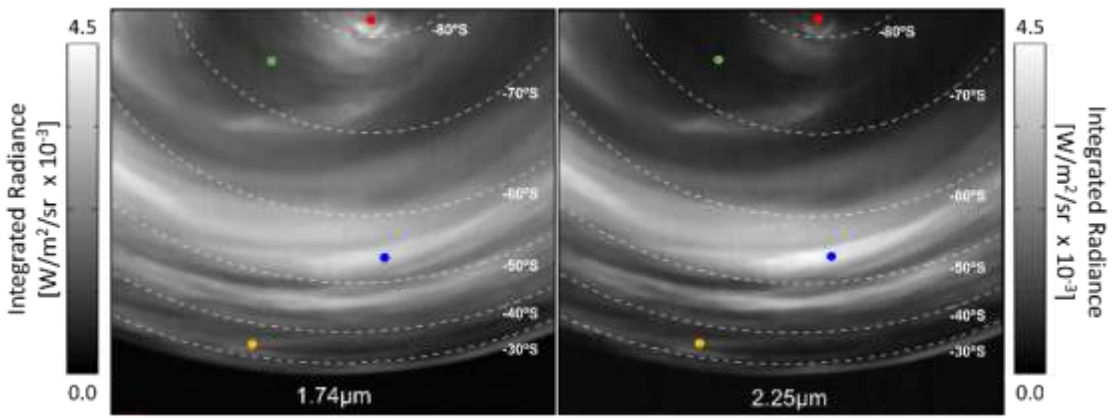

Figure 4. Comparison of the integrated radiance for the two wavelength bands in the same observation (VI0025_07, taken 15 May 2006 at 14:36 UTC) of the southern hemisphere of Venus. The coloured dots mark four representative latitude regions (Red = polar vortex; Green = cold collar, Blue = mid-latitude, Orange = low-latitude).

In Figure 4 we can distinguish similar cloud features appearing at both wavelengths, although with slightly different contrast. Note in particular the darker polar region from the cold collar to the pole (latitudes from -70ºS to -80ºS) seen around 2.25µm (right) with respect to the 1.74µm (left). These images are shown as obtained directly from the calibrated data file in spectral radiance, without any projection or limb darkening correction. The input spectral radiance obtained from the calibrated data is integrated over the length of the windows and multiplied by the spectral width of the band ($10^{-2}$µm) to convert to integrated radiance units (W/m$^2$/sr).

The main data selection parameters used for the integrated radiance of both atmospheric windows around 1.74µm and 2.25µm are listed in Table 1.

| *Atmospheric window* | *1.74µm* | *2.25µm* |
|---|---|---|
| *Wavelength range for radiance integration* | [1.68µm-1.78µm] (+/-10nm) bands 70-80 | [2.18µm-2.30µm] (+/-10nm) bands 122-135 |
| *Wavelengths used as reference for continuum subtraction* | [1.67µm, 1.79µm] (+/-10nm), bands 69,81 | [2.17µm,2.51µm] (+/-10nm) bands 121-157 |
| *Incidence angle filter* | >100° | >100° |
| *Emission angle filter* | <80º (to avoid limb observations) | <80º (to avoid limb observations) |
| *Limb darkening correction* | R / (0.34 + 0.66 · cos(EA) ) R: Radiance EA: Emission Angle | R / (0.23 + 0.77 · cos(EA) ) R: Radiance EA: Emission Angle |
| *Exposure time filter* | >0.1 seconds (exclude dayside observations) | >0.1 seconds (exclude dayside observations) |
| *Total observations used* | 3133 | 3133 |

Table 1. List of data selection parameters used to build the mean global maps of integrated radiance in the two main infrared atmospheric windows

The selection of wavelengths is done using a different range of bands for each case. The variability in the wavelength range (+/-10nm) needs to be considered in each individual observation to correctly accommodate the temperature differences that may affect the spectral registration in different measurements [Cardesín-Moinelo 2010]. In any case the selection of bands has been done carefully with enough margins to accommodate for temperature variations between or within observations. The extreme bands outside each window have been used to subtract the continuum and remove the effect of background thermal contribution, avoiding any additional potential problems related to temperature.

In our study, we have used all the available data but we have used a filter of incidence angle to make sure we only consider data from the nightside. The filter value used (>100º) includes a margin of 10 degrees over the terminator (90º) in order to exclude not only the data from the dayside but also any possible radiance scattered close to the terminator or the poles. For this reason we focus our analysis within +/-5h in local time and +/-75º in latitude.

The limb darkening correction has been performed to the data following the functions and average coefficients given by [Longobardo 2012] and applied to the data at all latitudes over the planet. In order to avoid problems with very high emission angles, an extra filter is applied to take into account only the data below 80º, that is ignoring all data close to the limb (emission angle 90º).

### 2.3. DATA SELECTION FOR THERMAL BRIGHTNESS ANALYSIS

The two thermal regions of interest, 3.8µm and 5.0µm, are shown in Figure 5. VIRTIS observations at these wavelengths, shown in Figure 6 usually show a strong correlation as the radiance at these bands depends almost purely on the thermal emission from the cloud tops at about the same level.

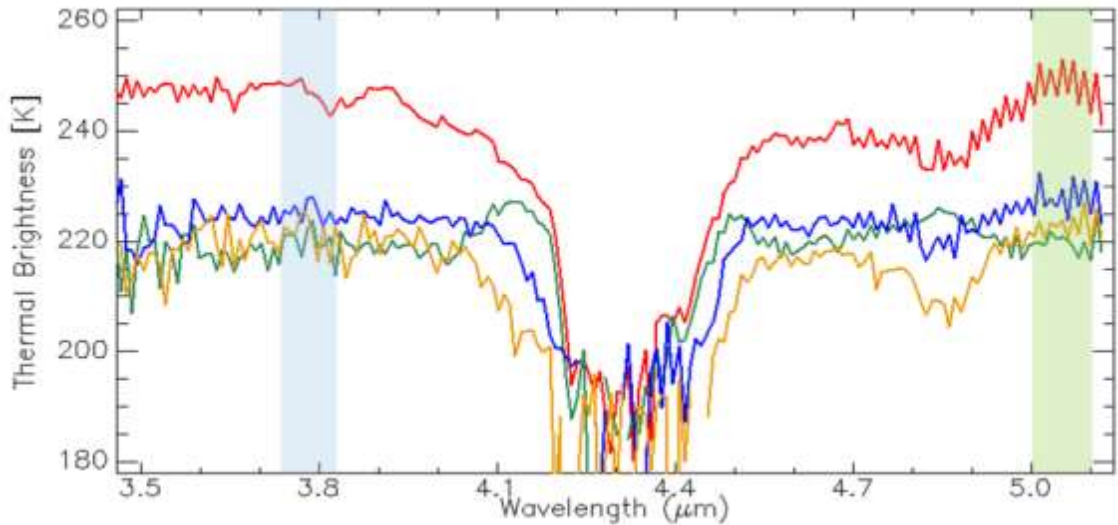

Figure 5. Typical VIRTIS-M infrared spectrum in the nightside of Venus between 3.5 and 5.1μm showing the main thermal regions used in this study, around 3.8 (marked in blue) and 5.0μm (marked in green). The different coloured plots show typical profiles in the thermal brightness as measured at various reference points of the nightside atmosphere: Red = polar vortex; Green = cold collar, Blue = mid-latitude, Orange = low-latitude.

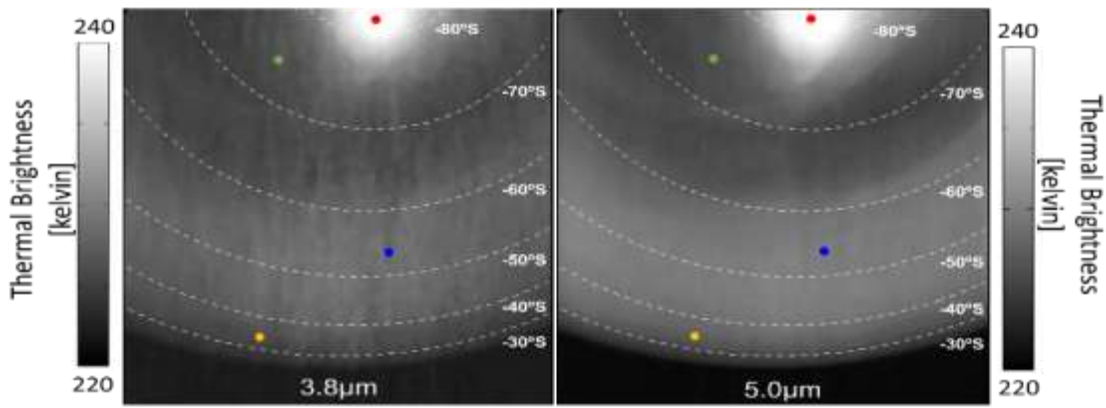

Figure 6. Comparison of the thermal brightness observed in the two wavelength bands in the same observation (VI0025_07, taken 15 May 2006 at 14:36 UTC) of the southern hemisphere of Venus. The coloured dots mark four representative latitude regions (Red = polar vortex; Green = cold collar, Blue = mid-latitude, Orange = low-latitude).

We can see similar features from the 3.8μm (Figure 6 left) and the 5.0μm (Figure 6 right), although with slightly different contrast. The images are shown as obtained directly from the calibrated data file without any projection or correction, only averaged and converted to thermal brightness in kelvin.

The main data selection parameters used for the thermal brightness both at 3.8µm and 5.0µm are reported in Table 2.

| *Thermal emission region* | *3.8µm* | *5.0µm* |
|---|---|---|
| *Wavelength range for thermal emission* | [3.73µm-3.82µm] (+/-10nm), bands: 287-296 | [5.01µm-5.10µm] (+/-10nm), bands: 421-431* <br><br> *Even bands are ignored to avoid "odd-even effect" [Cardesín-Moinelo, 2010] |
| *Mean Thermal Brightness* | All bands are converted into Thermal Brightness and then averaged together | All bands are converted into Thermal Brightness and then averaged together |
| *Incidence angle filter* | >100° | >100° |
| *Emission angle filter* | <60º (to reduce limb darkening) | <60º (to reduce limb darkening) |
| *Limb darkening correction* | R / (0.13 + 0.87 · cos(EA) ) <br> R: Radiance <br> EA: Emission Angle | R / (0.20 + 0.80 · cos(EA) ) <br> R: Radiance <br> EA: Emission Angle |
| *Exposure time filter* | >0.1 seconds (exclude dayside observations) | 0.1-2 seconds (to avoid data saturation) |
| *Total observations used* | 3034 | 1524 |

Table 2. List of data selection parameters used to build the mean global maps of thermal brightness in the two main thermal emission regions.

The selection of wavelengths is done using a different range of bands for each case. Again the variability in the wavelength range (+/-10nm) is mentioned to accommodate for possible variations in the spectral registration due to temperature differences.

The input data are obtained directly from the VIRTIS calibrated cube radiance, converted into Thermal Brightness [K], supposing an emissivity equal to 1.

Again we have used all the available data and applied the same filter to the incidence angle (>100º) to consider only nightside data and avoid any possible scattered radiance from the terminator or poles, focusing our analysis within +/-5h in local time and +/-75º in latitude.

The limb darkening correction for the thermal emission coming from the upper clouds is more complex than in the case of the lower clouds and has several limitations. The available methods proposed in [Longobardo 2012, Grinspoon 1993 and Roos 1993] are valid only in a restricted range of emission angles and the proposed coefficients are highly dependent on latitude. In order to address this problem we have first reduced the filtering threshold for emission angles below 60º, adding extra margin to avoid deviations at higher angles. Then we have performed tests with many correction coefficients and we have selected the values that, when applied to the whole planet, produced the optimal results based on consistency in comparison with recent thermal studies by [Grassi 2008, 2010, 2015], [Migliorini 2012], [Haus 2013] and the review by [Limaye 2018b]. In the case of 3.8µm we have selected the coefficients (0.13/0.87), which were already proposed both by [Longobardo 2012 and Grinspoon 1993] for the 3.72µm wavelength. For 5.0µm we have selected the coefficients (0.20/0.80), originally proposed by [Longobardo 2012] for the 4.0µm wavelength at low latitudes.

As previously mentioned, the thermal radiance around 5µm is heavily affected by the instrument thermal noise that saturates at high exposure times, so only the short exposure observations (<0.36s) can be used for this wavelength, with some important implications. First the number of observations at 5µm is much lower than for other wavelengths, so the coverage is reduced significantly, especially in the northern hemisphere where the coverage is already limited by the observing geometry. Secondly, the low exposure observations are heavily affected by an instrumental anomaly called "odd-even effect", caused by small differences in the responsivity of the infrared sensor [Cardesín-Moinelo, 2010]. This affects especially the wavelength region around 5µm, producing a "sawtooth" spectral irregularity, well visible in Figure **5**. In order to limit this effect, we have used only odd bands, ignoring all even bands in the 5µm region. Note this instrumental effect does not affect other wavelength regions, where the overall effect can be ignored because the exposure times can be longer as well.

## 3. RESULTS: INTEGRATED RADIANCE IN THE ATMOSPHERIC WINDOWS

### 3.1. GLOBAL MAPS OF THE INFRARED ATMOSPHERIC WINDOWS

The mean global maps distributed along Latitude and Local Time are shown in Figure 7, for the integrated radiance in the window centred at 1.74µm and Figure 8 for the window centred at 2.25µm. These maps have been produced using a grid of 5º in latitude and 30minutes in local time. To facilitate the comparison between these two bands we have also produced a map with the ratio between 1.74/2.25µm. This ratio is meant to facilitate the comparison between the two bands and it also has the advantage that it can provide an indirect estimation of the particle size as described by previous studies [Carlson 1993] [McGouldrick 2008, Wilson 2008, Haus 2013]. Note however that a detailed analysis of the particle size is out of the scope of this study and we only provide here a basic estimation of the relative size spatial distribution assuming that the averaged region has a uniform cloud cover. This is a very simplistic approach with many limitations but it can still be useful to identify the main global trends in the atmosphere that we will discuss in the following sections compared to previous studies.

Maps for both infrared windows are well correlated and show basically the same global characteristics, in particular the increase of radiance at mid-latitudes, with a relatively good symmetry in both hemispheres. The more patchy distribution in the north is likely caused by the more limited coverage as previously mentioned. The increase of radiance translates into a more transparent clouds deck in the same region, very likely associated to downwelling due to the meridional cell. On the contrary, the low radiance is where the clouds thickness is maximum, like in the equatorial region and even more pronounced in both polar regions.

Note that nightside data has been filtered with an incidence angle threshold >100º, including a margin to reduce any possible straylight or scattering from the dayside terminator. Note in particular that the data around both poles (>80º and <-80º) and around the terminators is outside this threshold and therefore it is not shown to avoid misinterpretations.

The general features and trends can be similarly seen in both spectral regions but with small and important differences in the details. For example, there is a different contrast especially at high latitudes, due to the cold collar around the polar region [Piccioni 2007, Carlson 1993] and the increase of abundance of large particles within the polar vortex, which enhances a lot the contrast in the 2.25µm with respect to the 1.74µm [Wilson 2008].

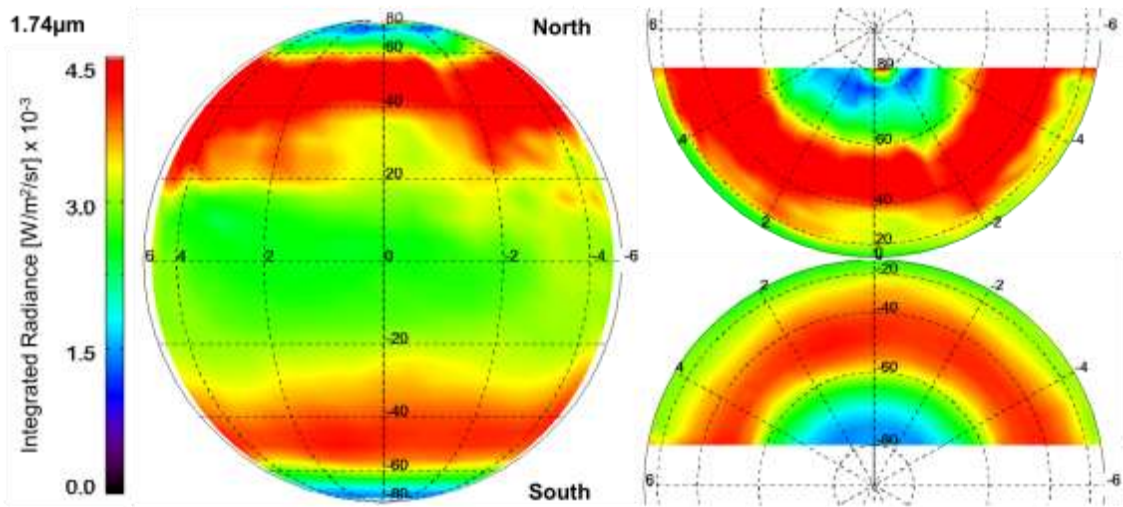

Figure 7. Map of integrated radiance in the window centred at 1.74µm, shown in an orthographic projection of the Venus nightside hemisphere, distributed along Latitude (north up, south down) and Local Time. The left projection is centred at the anti-solar point (midnight, local time 00h). The right projections are centred at the poles.

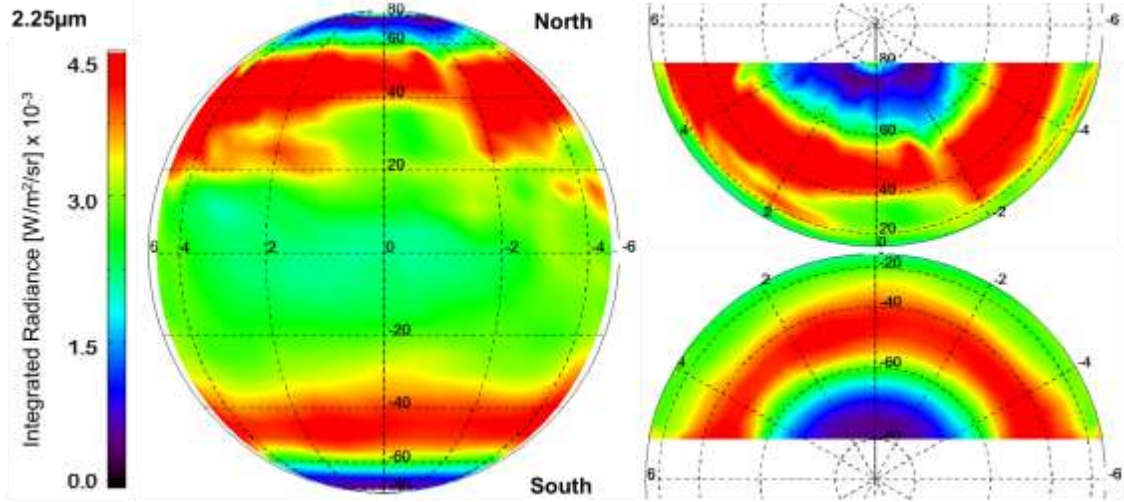

Figure 8. Map of integrated radiance in the window centred at 2.25μm, shown in an orthographic projection of the Venus nightside hemisphere, distributed along Latitude (north up, south down) and Local Time. The left projection is centred at the anti-solar point (midnight, Local time 00h). The right projections are centred at the poles.

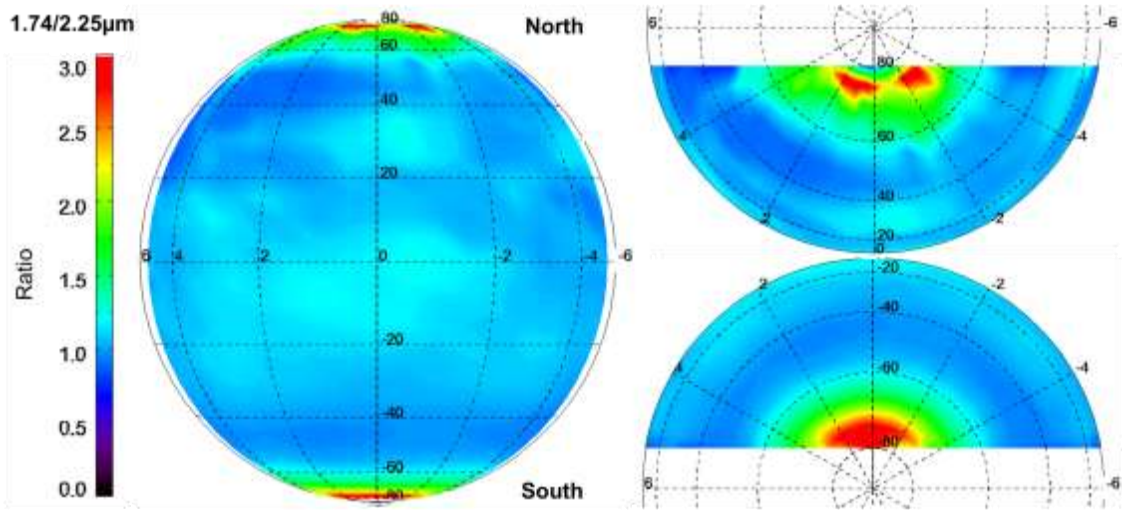

Figure 9. Ratio between the integrated radiances measured at 1.74μm and 2.25μm. Shown in an orthographic projection of the Venus nightside hemisphere, distributed along Latitude (north up, south down) and Local Time. The left projection is centred at the anti-solar point (midnight, Local time 00h). The right projections are centred at the poles.

## 3.2. DETAILED INFRARED ATMOSPHERIC WINDOW PROFILES

We now show a more detailed trend distribution of the integrated radiances, with mean profiles with respect to latitude and local time, as in Figure 10.

The uncertainty of the spectral radiance measurements by VIRTIS-M-IR is below $10^{-3}$ W/m$^2$/sr/µm [Piccioni 2008, Cardesin-Moinelo 2010] with a bandwidth of ~10nm. This translates into an integrated radiance below $10^{-5}$ W/m$^2$/sr, therefore random measurement noise is negligible for our study and no error bars are shown in the profiles. Other possible sources of non-random errors, in particular systematic offsets or potential errors in the data calibration or the data processing may still be present in the results. In any case, the main source of error in our profiles is driven by the large fluctuations of the atmosphere. This is well visible in the latitude profiles for the northern hemisphere with very large variations caused by the lower coverage in the north, which make the atmospheric fluctuations more evident with respect to the smoothed, averaged values of the southern hemisphere. Also the scattered radiation coming from the strong illuminated day side affects all the averaged measurements near the terminator and the poles and therefore the data close to the terminator is not used in the analysis.

These profiles show the general latitudinal trend with low radiance (dense lower clouds) at the equator and the poles, and a relatively higher radiance (thinner lower clouds) at mid latitudes, around +/-45°. Interestingly, the variability from the relative maximum peaks at mid latitudes and the minimum at the equator is about the same for both wavelengths. Indeed the latitudinal variation of the ratio is very flat in mid-to-low latitudes, indicative of a very similar opacity and particle size in these regions. This is very different at high latitudes around the minimum radiance values at +/-75º, where 2.25µm integrated radiance gets 2.5~3 times dimmer than at 1.74µm, indicating a large particle size, very symmetrical at both hemispheres despite the coverage limitations. Another interesting feature is the slight asymmetry of the peaks at mid latitudes, with higher values in the north (+75º) than in the south (-75º).

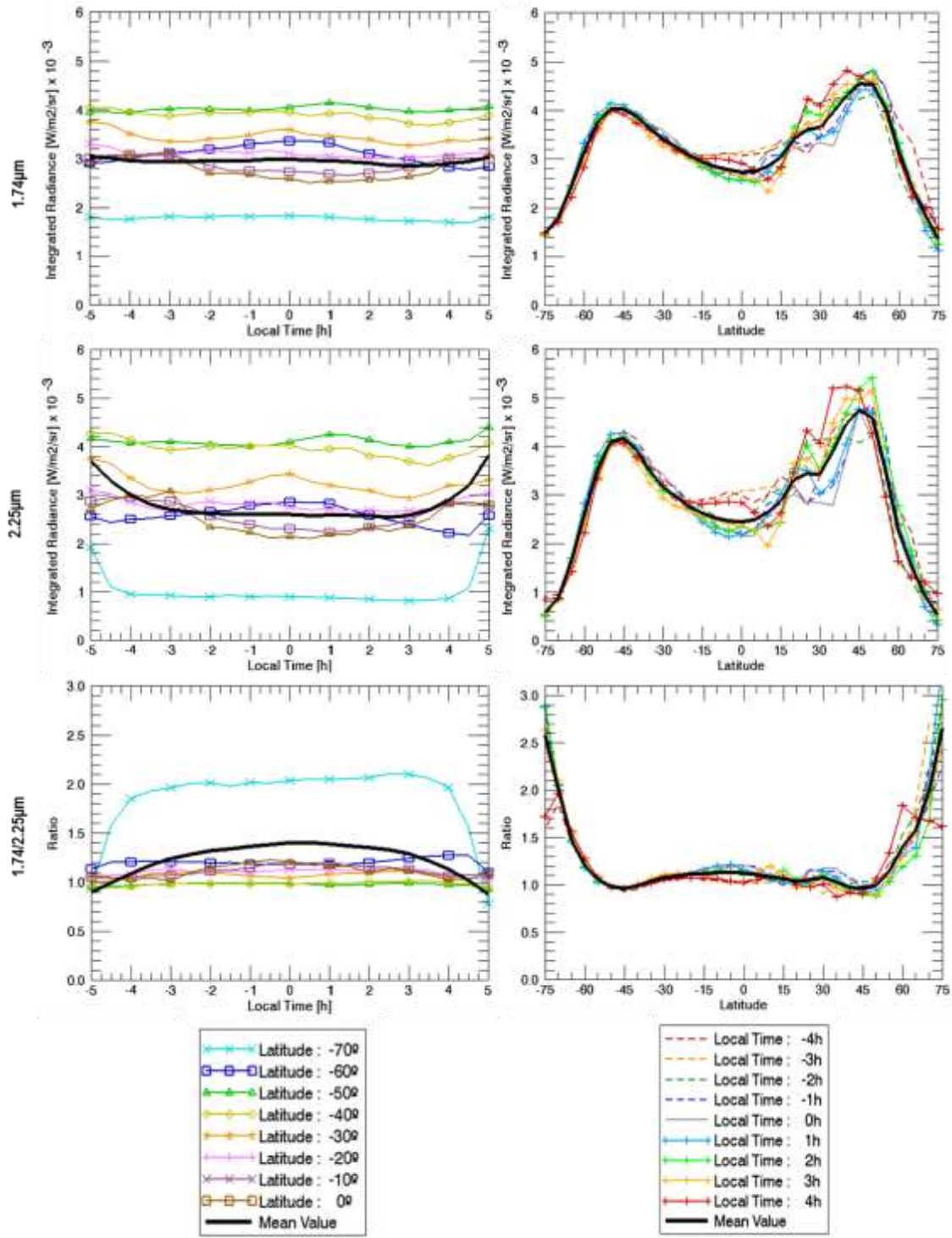

Figure 10. Radiance profiles at 1.74μm (top), 2.25μm (middle) and ratio between them 1.74/2.25μm (bottom) as a function of local time (left) and latitude (right). The regions close to the terminator and polar caps are omitted since they are affected by scattered radiation.

The mean radiance distribution in local time can also be seen in Figure 10. The steep increase of radiance on the left and right sides close to the terminator is explained again by scattering contamination from the day side, which affects the weak signal from the night side. The central part, from -4h to +4h, is relatively flat with the exception of sub-tropical (+/-30º) and sub-polar (+/-60º) latitude regions, which show an interesting peak around midnight. The higher contrast in radiance at 2.25µm with respect to 1.74µm between the polar and equatorial regions is obviously present also here.

### 3.3. LONGITUGINAL VIEWS OF INFRARED ATMOSPHERIC WINDOW

We will now present the nightside integrated radiance distributed in a longitudinal view for the southern and northern hemispheres. The maps are shown in Figure 11 and Figure 12 for the bands 1.74 and 2.25µm respectively. Both bands are projected spatially with respect to Latitude and Longitude using a grid of 5x5 degrees and centred at each pole. This perspective visually enhances the global extension of the high radiance at mid-latitudes, as well as the difference in relative radiance near the poles between the two atmospheric windows.

A preliminary version of these polar views was previously presented for the southern hemisphere by [Cardesín-Moinelo 2008] and briefly discussed by [Titov 2018]. These new maps have been reprocessed covering both hemispheres and include various updates, in particular we now use better emission angle corrections based on [Longobardo 2012], new grid (5x5 degrees in latitude/longitude) and some other adjustments and filtering of incorrect and meaningless data.

All these improvements provide better estimations of the mean radiance values and allow for a better comparison of both wavelengths. We can remark again the global latitudinal trends that are common to both wavelengths, and the much larger depletion region around the pole for the 2.25µm radiance with respect to the 1.74µm one.

We note here that the longitudinal maps show some irregularities and asymmetries with respect to longitude, but any assumption regarding surface-cloud interactions is out of the scope of this study and in any case not obvious from our results. Several other investigations have already discussed interesting dependencies between the surface and the clouds, wind variations localized around certain regions as identified with Venus

Express data by [Bertaux 2016] and [Peralta 2017b], and the recent studies based on Akatsuki bow-shape features by [Fukuhara 2017, Kouyama 2017], complemented also with modelling analysis by [Kashimura 2019].

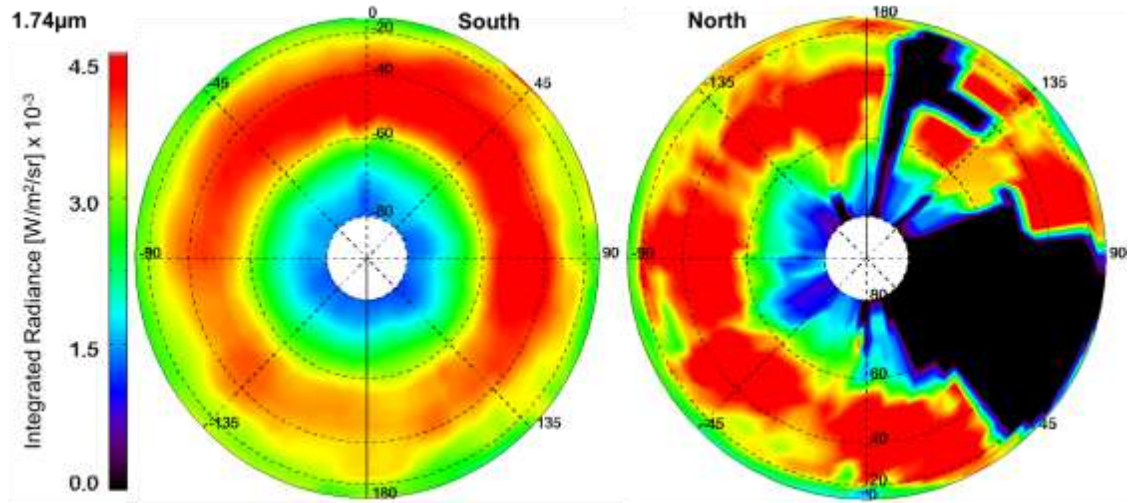

Figure 11. Mean radiance integrated in the window at 1.74µm, shown in an orthographic projection of the Venus southern (left) and northern (right) hemispheres, distributed along Latitude (pole in the centre) and Longitude (increasing eastward).

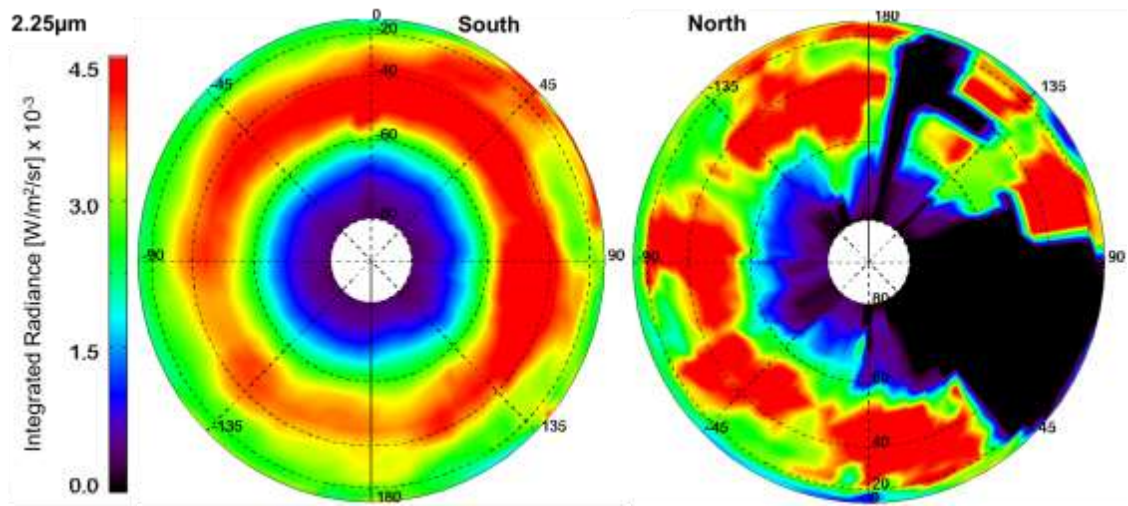

Figure 12. Mean radiance integrated in the window centred around 2.25µm, shown in an orthographic projection of the Venus southern (left) and northern (right) hemispheres, distributed along Latitude (pole in the centre) and Longitude (increasing eastward).

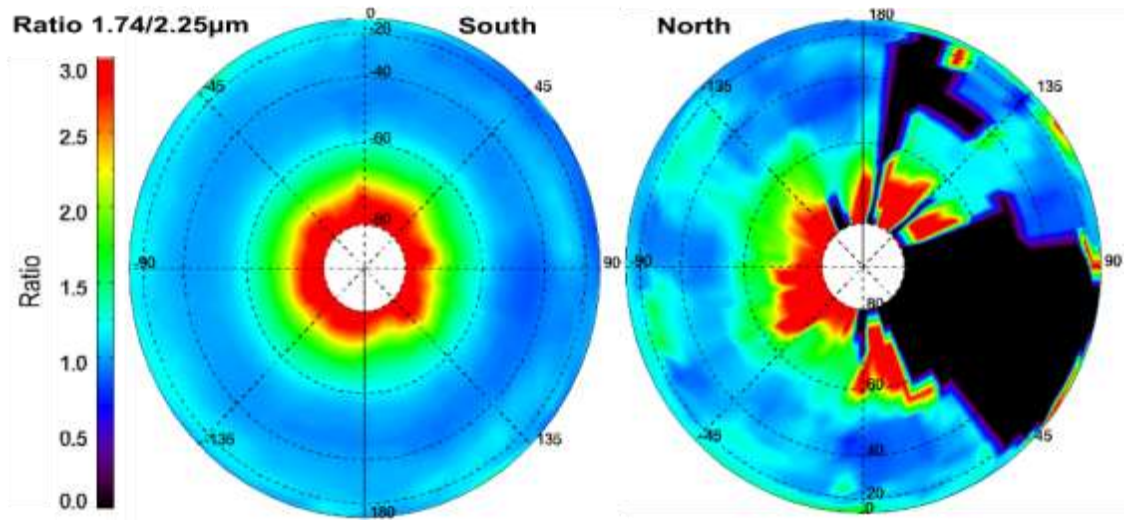

Figure 13. Ratio between the radiances measured at 1.74μm and 2.25μm. Shown in an orthographic projection of the Venus southern (left) and northern (right) hemispheres, distributed along Latitude (pole in the centre) and Longitude (increasing eastward).

## *4.* RESULTS: BRIGHTNESS TEMPERATURE OF INFRARED THERMAL EMISSION

### 4.1. GLOBAL MAPS OF INFRARED THERMAL EMISSION

We will now present the global maps that have been done for the thermal brightness. These maps are shown in Figure 14 for 3.8µm, and Figure 15 for 5.0µm, with orthographic projections centred at the anti-solar point and using a grid of 5 degrees in latitude and 30minutes in local time.

Note again that nightside data has been filtered with an incidence angle threshold of >100º, including a margin to reduce any possible straylight or scattering from the dayside terminator. Data around both poles (>80º and <-80º) and around the terminators is outside this threshold and not shown in the maps.

The 5.0µm map has large gaps especially in the northern hemisphere, as this wavelength region was saturated for long exposure times and therefore it includes a smaller number of observations with respect to the 3.8µm map. The two maps have a very similar appearance, also quantitatively, and both show consistently a nice distribution of the cloud tops temperature of the full Venus night side.

Very interesting features are evident in the maps. First the uniform temperatures at low latitudes, with no evident variability in latitude or local time. Then at mid-to-high latitudes we start seeing the temperatures decreasing smoothly from the evening to the morning side in both hemispheres, especially the collar region centred at +/-60° around both poles. This clear symmetry in latitude and local time for both hemispheres is consistent with the behaviour seen by [Grassi 2010, Grassi 2014, Migliorini 2012] and the modelling simulations by [Haus 2013, Garate-Lopez 2018].

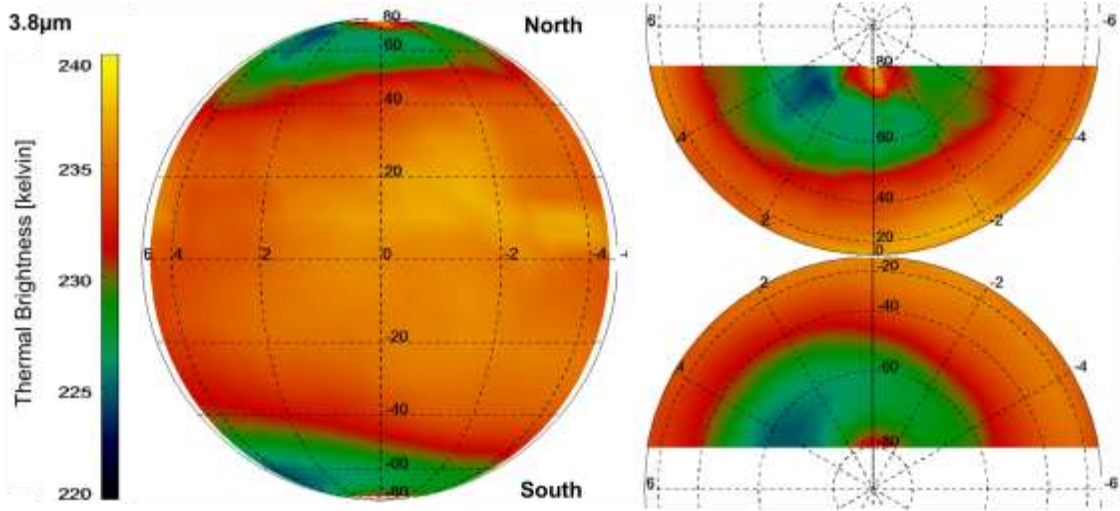

Figure 14. Thermal Brightness mean map of Venus at 3.8μm, shown in Orthographic projection, distributed along Latitude (north up, south down) and Local Time. The left projection is centred at the anti-solar point (Midnight, Local Time=00h). The right projections are centred at the poles.

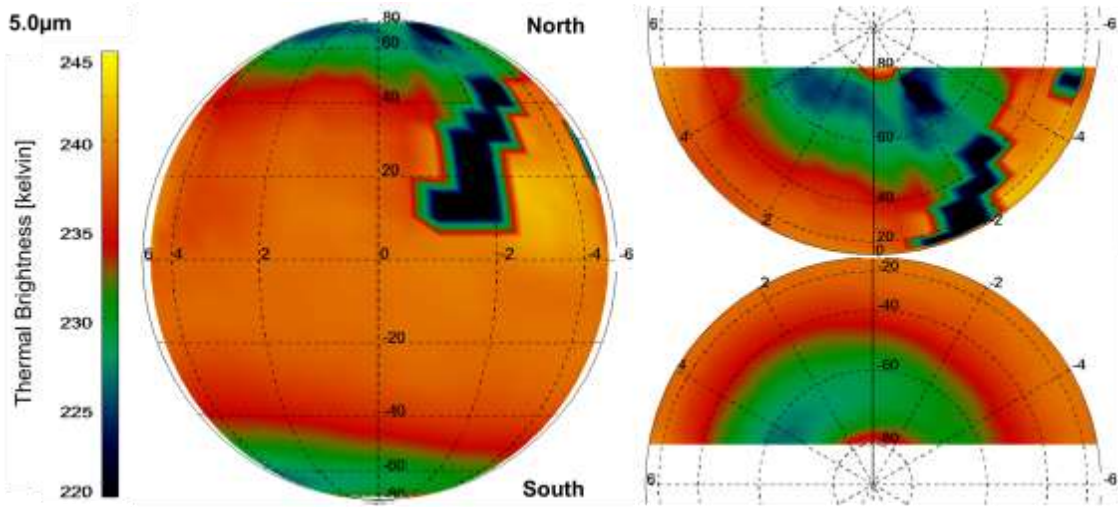

Figure 15. Thermal Brightness mean map of Venus at 5.0μm, shown in Orthographic projection, distributed along Latitude (north up, south down) and Local Time. The left projection is centred at the anti-solar point (Midnight, Local Time=00h). The right projections are centred at the poles.

## 4.2. DETAILED THERMAL BRIGHTNESS PROFILES

The mean thermal brightness profiles with respect to latitude and local time are reported in Figure 16. In particular, we show the mean profiles at 3.8µm and 5.0µm at different latitudes versus local time and different local times versus latitude. The profiles of both bands are qualitatively and also quantitatively consistent with each other with very small differences. As previously mentioned, the uncertainty of the input spectral radiance values measured by VIRTIS-M-IR is negligible compared to the large fluctuations of the atmosphere, therefore error bars are not shown in these profiles. We remark that both bands have the minima in the cold collar at about +/-60~70° of latitude and the relative maximum around the equator. Regarding the differences between both bands, we note a lower thermal brightness of the northern cold collar seen at 5.0µm and many other small differences in the northern hemisphere, which can all be explained by the lower coverage.

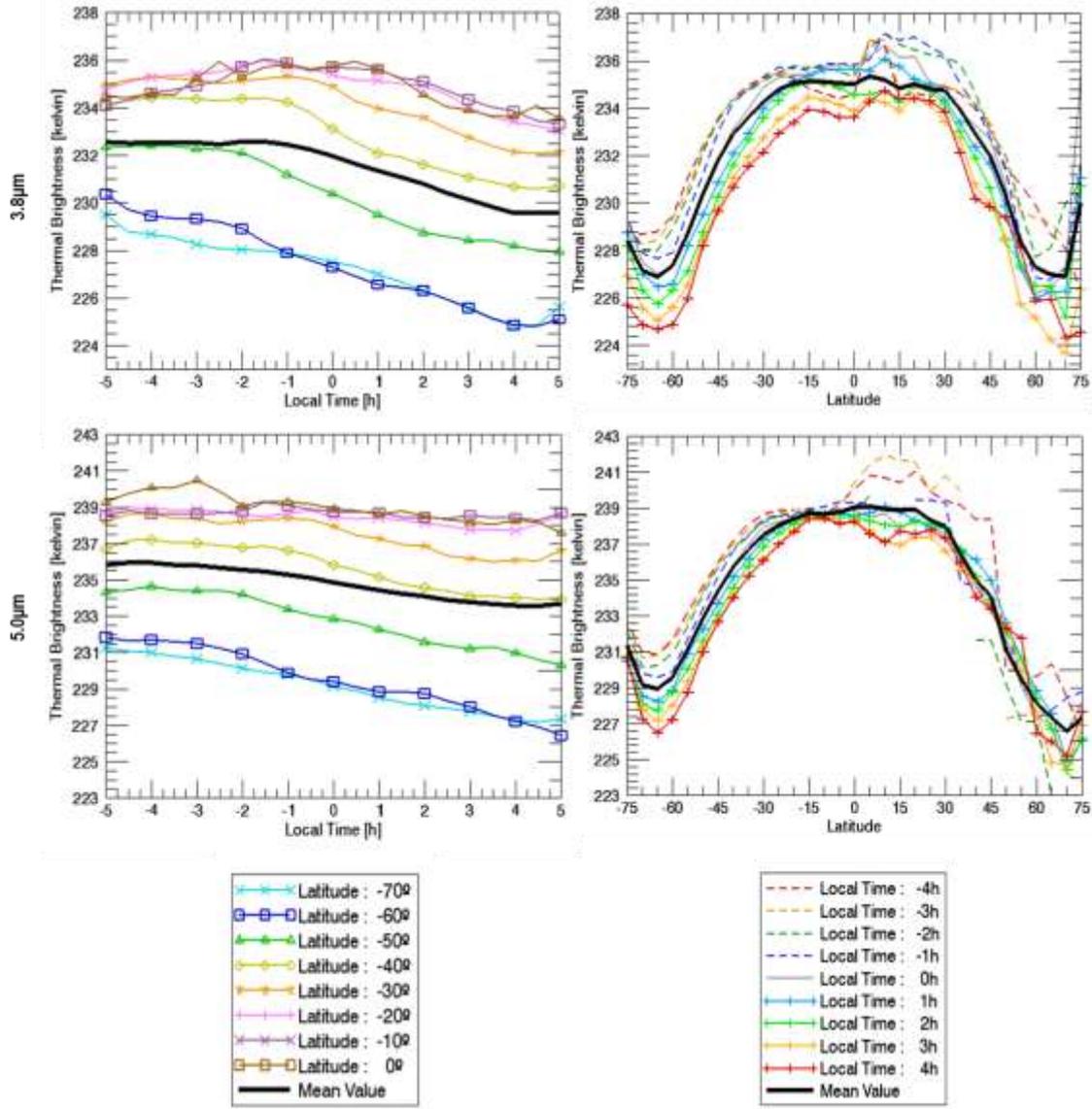

Figure 16. Profiles of Thermal Brightness at 3.8µm (top), 5.0µm (bottom) plotted with respect to local time (left) and latitude (right). The regions close to the terminator and the polar caps are omitted since they are affected by scattered radiation.

The local time profiles shown in Figure 16 are also consistent for the two wavelengths. The more evident behaviour of these profiles are the colder temperatures seen in the morning side (right) with respect to the evening side (left) with a very smooth and regular trend from one side to the other, with a difference of about 3 to 5K in average. It is also remarkable also the fact that the largest dependency with local time corresponds to high latitudes where the Sun's elevation is very low and the average rotation period of Venus atmosphere is smaller with respect to lower latitudes, therefore the differences between day and night at high latitudes would not be expected to be so strong.

### 4.3. LONGITUDINAL VIEWS OF THERMAL BRIGHTNESS

We will now present the nightside thermal brightness distributed in a longitudinal view for the southern and northern hemispheres. These maps are shown in Figure 17 for 3.8μm and Figure 18 for 5.0μm. All maps are projected spatially with respect to latitude and longitude using a grid of 5x5 degrees and centred at each pole.

These polar views illustrate the uniform temperature brightness at low-to-mid latitudes and the sharp decrease in the cold collar around both poles. We can note an interesting asymmetry around the south pole with the cold collar not being uniform along the longitudes, visible in both bands. We cannot take any conclusions regarding this potential shift as this could be a long term evolution of the meridional displacement of the cold collar, not enough covered by the Venus Express data range with time, and the asymmetry of the data coverage in the eastern/western hemispheres as visible in Figure 2. As previously mentioned, any further assumption regarding surface-cloud interactions is out of the scope of this work and one should refer to other relevant studies mentioned in Section 3.3.

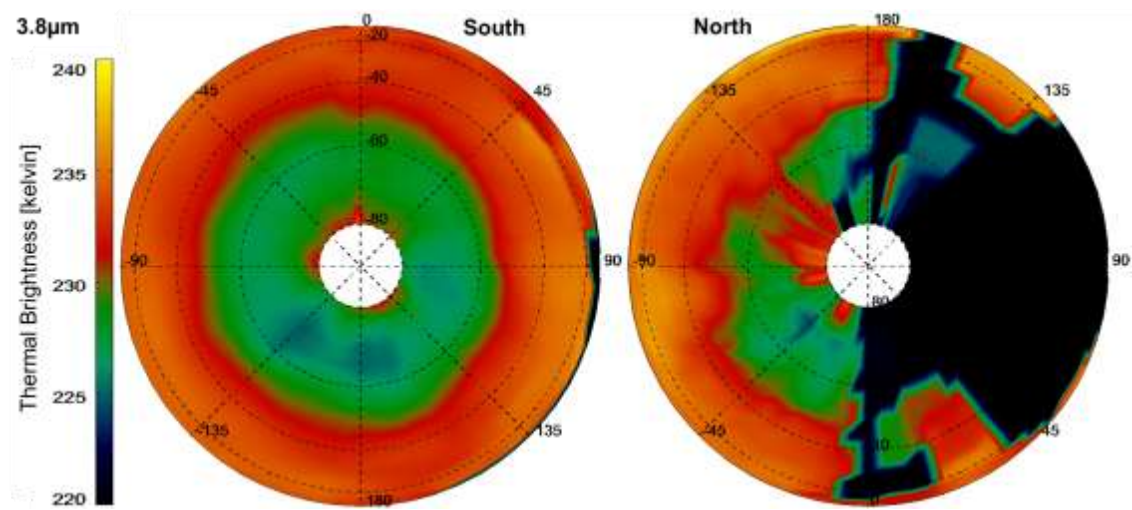

Figure 17. Map of mean thermal brightness at 3.8μm, shown in an orthographic projection of the Venus southern (left) and northern (right) hemispheres, distributed along Latitude (south pole in the centre) and Longitude (increasing eastward).

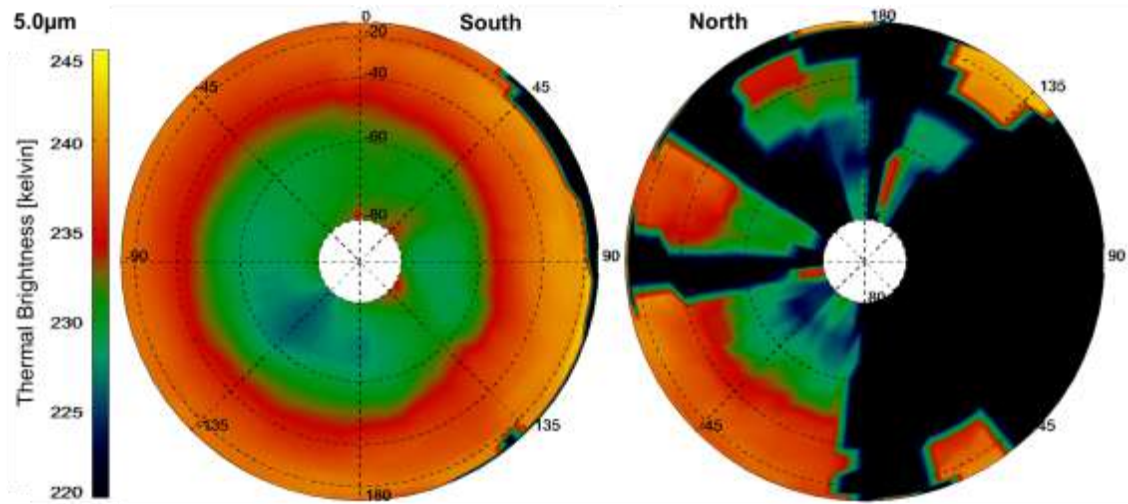

Figure 18. Map of mean thermal brightness at 5.0μm, shown in an orthographic projection of the Venus southern (left) and northern (right) hemispheres, distributed along Latitude (south pole in the centre) and Longitude (increasing eastward).

## 5. DISCUSSION

All the maps indicate generally four very well-defined regions: the equatorial belt, the mid-latitude regions, the cold collars and the polar regions.

### 5.1. EQUATORIAL REGION (LATITUDES 0-30º)

The maps show an equatorial belt with relatively high cloud opacity. The equatorial belt of Venus is dominated by generally thick clouds [Titov 2018]. The low integrated radiance observed all around the equator is explained by the thermal radiation coming from the layers below the lower clouds, getting absorbed by the thick equatorial clouds, causing lower values in average at the equator. Both atmospheric windows show a very similar profile and indeed the band ratio shows no significant variability in this region, implying a uniform spatial distribution of the particle sizes with local time or latitude around the equator, consistent with [Wilson 2008].

The thermal emission maps show a uniform brightness temperature at the cloud tops, with very high values and low variability around the equator consistent with the intersection of the expected thermal profiles [Limaye 2018b]. This is well in agreement with the smooth symmetric temperature profiles reported by [Migliorini 2012] and [Haus 2013] at these latitudes, and it is also consistent with the altimetry of the clouds retrieved in [Ignatiev 2009]. Regarding the local time variations in this low latitude region, the profiles are also very flat and there are very small variations between evening and morning hours, except those attributed to the limited coverage at these latitudes.

### 5.2. MID-LATITUDE REGION (LATITUDES 30-60º)

The mid latitude region, centred between -30º and -60º of latitude, shows values of integrated radiance higher than the surroundings in average. The lower opacity of these thinner clouds are due to a more regular laminar flow motion described in [Sanchez-Lavega 2017] and the deep radiance penetrates the clouds toward the outer space relatively easier in this region [Titov 2018]. This is consistent with previous observations from Earth, see [Tavenner 2008], and results from other missions as summarized by [Peralta 2018]. The mid latitude is then the region where the highest dissipation of the surface or close to surface heat can take place, an important point for the differential (meridional) and absolute thermal balance of the planet [Limaye 2018b].

This region is very evident in the integrated radiance maps (Figure 7 and Figure 8) and profiles (Figure 10) getting to the maximum values around +/-45º latitude, with values exceeding $4\times10^{-3}$ W/m$^2$/sr for both the 1.74µm and 2.25µm bands, with no significant variation in the ratio between them. This is a transition zone where the atmospheric flow is more laminar and where there is a "clearing" effect on the clouds from the boundary of the cold collar and polar vortex dynamics. In this respect, [Kashimura 2019] demonstrates the morphological similarity between the night-side bright streaks and down-welling region in high-resolution numerical simulation. Our maps of integrated radiance maps show a symmetric behaviour in both hemispheres. The maximum values are slightly asymmetric, higher in the north in both wavelengths (1.74µm and 2.25µm) and is consistent with recent Akatsuki IR2 images at 2.3µm that showed differences between both hemispheres at low latitudes [Peralta 2018] and [Peralta 2019], although these reports were not systematic analyses of long term observations like the results presented in this work.

We may also discuss here the interesting bulge in radiance observed around midnight in the sub-tropical (+/-30º) and sub-polar (+/-60º) latitudes, shown in Figure 10. This higher values at the boundaries of the mid-latitude region appear consistently in both bands, 1.74µm and 2.25µm, with no clear variation in the ratio between them, and it is very reliable since the coverage in the southern hemisphere is very good with nearly a thousand observations. On the other hand this is not visible in the northern hemisphere, although this could be a consequence of the lower coverage and higher variability. This interesting increase in the radiance around midnight can be explained by an enlargement in the meridional direction of the mid-latitude thin cloud belt towards the equator. We remark that this potential expansion has not been reported in previous studies and shall be studied in future simulations with General Circulation Models.

It is worth mentioning that despite the great variation in the integrated radiance from low to mid-latitudes, the variations of band ratio stay almost flat and symmetric between 60ºS and 60ºN. In general this is consistent with [Wilson 2008] and [Barstow 2012] that reported that latitudinal variations of cloud thickness are not directly correlated with the particle size in this region. When looking at our detailed profiles, we can identify a small decrease of the ratio by 10~15% around +/-60º with respect to the equator, and again symmetric in both hemispheres. This decrease in the ratio is only a potential indication of smaller particle size, but is in good agreement with the thinner mid latitude clouds reported by [Haus 2013].

Regarding the local time dependencies, the integrated radiances do not present significant variations and all the mid latitude range has similar values. On the contrary, the thermal brightness profiles (Figure 16) show a clear trend of the cloud top temperatures decreasing with local time at these latitudes (see for example the local time profile at -50º latitude in both bands), decreasing by about 2K from the evening terminator towards midnight and then decreasing furthermore up to 5K towards the morning terminator. This is in line with the local time variations described for the southern hemisphere by [Grassi 2010 and 2014] and our results show a good symmetry with the northern hemisphere, especially at 3.8µm (Figure 14). The causes of this behaviour and the correlation with thermal tides have recently been investigated with modelling simulations (see [Ando 2016] and references therein). Also recent observations with Akatsuki/LIR data, reported by [Kouyama 2019], show the global structure of the thermal tides at the cloud top level, and are consistent with the local variation of temperature obtained from our VIRTIS thermal emission data.

The distribution of radiance and thermal brightness in the longitudinal maps (Figure 11, Figure 12, Figure 17 and Figure 18) offer a good illustration of the global extension of the thin clouds at mid-latitudes in both hemispheres and the sharp decrease, both in radiance and in cloud top temperatures, when and entering the cold collar around ~60° latitude.

### 5.3. COLD COLLAR REGION (LATITUDES 60-75º)

The convection-driven Hadley cell extends up to about 60° latitude in both hemispheres moving with a global trend from the equator to higher latitudes, feeding a mid-latitude jet at its poleward extreme, inside which there is a circumpolar belt characterized by remarkably low temperatures and dense high clouds. This is the so-called "cold collar" region [Taylor 1980, Piccioni 2007, Titov 2018, Sanchez-Lavega 2017], located between 60º and 75º latitude, which is clearly visible in all global maps for its lower radiance and brightness temperature.

The cold collar represents the important dynamical boundary between the polar region and the mid to low latitudes of Venus. The cold collar is suggested to be caused by a zonal jet produced by the mean meridional circulation, as simulated through a General

Circulation Model [Lee 2005]. More recently, the models by [Ando 2016, 2019] and [Garate-Lopez 2018] described the adiabatic expansion due to the Hadley cell and the polar vortex, with also vertical motion of air and the superposition of the zonally averaged basic field, thermal tides and transient waves.

The profiles we have obtained at 60ºS and 70ºS for both thermal emissions (Figure 16) suggest a local time variability of the temperature in the cold collar region with an evident decreasing trend from the evening towards morning hours. The brightness temperature reach the lowest values in these latitudes, decreasing from ~232 K at -5h down to ~226 K at +5h for the 5.0µm band, and from ~230 K to ~225 K for the 3.8µm band. This cooling trend was first reported in the northern hemisphere by [Taylor 1980] and then studied with detailed temperature fields for the southern hemisphere by [Grassi 2010, 2014]. Our results provide a very good latitudinal symmetry between northern and southern cold collars, especially evident at 3.8µm (Figure 14). A latitudinal symmetry was already described by [Migliorini 2012] and [Haus 2013] considering only averaged temperature fields without local time variation, and our results now confirm also that the local time profiles of both south and north cold collars have equivalent cooling trends from the evening towards the morning side. The causes of this local time dependence and the correlation with thermal tides are again described with modelling simulations (see [Ando 2016, 2019], [Garate-Lopez 2018] and references therein).

The cold collar region is also the region where the opacity of the clouds starts to increase with consequent lower radiance seen through the atmospheric windows. The integrated radiances for both atmospheric windows decrease from their maximum values at 60ºS latitude (both around ~4x$10^{-3}$ W/m$^2$/sr), to their minimum values at 75ºS, ~1.5x$10^{-3}$ W/m$^2$/sr in the 1.74µm band, and even lower to ~0.5x$10^{-3}$ W/m$^2$/sr in the 2.25µm band. These difference in the trend is clearly visible in the ratio latitudinal profile shown in Figure 9. The ratio is almost uniform and flat between 60ºS and 60ºN but increases sharply when entering the cold-collar region. This is caused by the different spatial distribution of size particles as explained by [Carlson 1993], with bigger particles depleting more the radiance at 2.25µm due to the higher imaginary part of the refractive index. Our maps of band ratio (Figure 9 and Figure 13) illustrate this particle size distribution and confirm that the concentration of bigger particles are symmetric in the southern and northern polar regions as reported by [Wilson 2008] and [Haus 2013].

In general we do not observe a dependence with local time inside the cold collar for the integrated radiance windows. In any case we can remark a 10% increase in the local time profile of the ratio at 70ºS (Figure 10, going from 1.9 at LT-4h to 2.1 at LT+4h). At this high latitude we cannot exclude a possible contamination by the scattered light from the terminator, but this could be an indication of a possible increase in the size of the particles from the evening towards the morning following the general atmospheric circulation and downwelling inside the cold collar [Titov 2008, Sanchez-Lavega 2017, Titov 2018, Garate-Lopez 2018, Lebonnois 2018].

### 5.4. POLAR REGION (LATITUDES 75-90º)

The study of this latitude region in our global maps is highly affected by the scattered radiation coming from the illuminated day side, so we avoid any interpretations close to the terminators and especially above 75º in latitude. What is evident is the big particle size and the higher thermal brightness seen in the polar regions up to at least 75º latitude in both hemispheres, consistent with the subsidence with clouds depression reported by [Titov 2008, Ignatiev 2009, Cottini 2012, Cottini 2015]. Most of the studies by Venus Express have focused in the Southern polar region but here we can remark the very good symmetry between the South and North polar regions, with no significant differences observed neither in the cloud top temperatures nor in the cloud opacity, except for the higher variability in the northern hemisphere due to lower coverage. This good symmetry is visible in most of the polar maps, and also in the latitudinal profiles in Figure 10 and Figure 16.

The polar regions are characterized by clouds thicker than in the cold collar, as reported by [Lee 2012] and possibly explained with mean meridional circulation responsible of cloud material transport [Imamura and Hashimoto 1998], yielding to higher temperature and adiabatic compression of the vortex dynamics, as discussed by [Garate-Lopez 2015]. Recent observations with IR2 on Akatsuki at latitudes >60° showed features darker than what observed at 2.02 µm, close to the cold collar borders, which are indicative of lower cloud top altitudes, due to the $CO_2$ absorption of reflected light [Limaye 2018a].

The polar vortex is only partially visible in our maps, and its study is out of the scope of this work. However, based on the good symmetry we observe between both polar regions, we can assume that the many attempts to understand the nature of the south polar vortex with VIRTIS data, starting by [Piccioni 2007], are also applicable to the North. We note in particular the study of the vortex motions discussed in great detail by [Luz 2011] and [Garate-Lopez 2013], its thermal structure investigated by [Grassi 2008] and [Garate-Lopez 2015] and its potential vorticity described by [Garate-Lopez 2016].

## 6. CONCLUSIONS

The large dataset of the VIRTIS instrument, operated from May 2006 to October 2008 on board the Venus Express mission, provides an extensive and unprecedented view of the Venus nightside, in particular in the near infrared atmospheric windows and thermal emission radiance. Despite the known large variability of the Venus atmosphere and clouds morphology, our mean global maps show a clear distribution of the different dynamical regions of the planet seen throughout their modification of the appearance of the clouds opacity and cloud tops thermal brightness.

A very good and comprehensive coverage is obtained for the southern hemisphere, while a more limited coverage is achieved in the North due to orbital constraints. However the general overall features observed appear symmetric on both hemispheres.

The maps show an equatorial belt with relatively high cloud opacity, spatially homogeneous particle size and a uniform brightness temperature at the cloud tops, in agreement with the smooth symmetric temperature profiles reported by [Migliorini 2012] and [Haus 2013] and also consistent with the altimetry of the clouds retrieved in [Ignatiev 2009].

The mid latitude region shows values of integrated radiance higher than the surroundings in average due to the lower opacity of the thinner clouds, more regular and with laminar flow motion as described in [Sanchez-Lavega 2017] and [Titov 2018]. This region also shows the smallest particle size compared to any other latitude range, confirming the results by [Haus 2013]. Here the integrated radiances do not present significant variations, but we can remark a potential enlargement of the southern mid-latitude belt towards the equator, not reported in previous studies. The thermal brightness show a clear trend of the cloud top temperatures decreasing with local time in both hemispheres.

In the cold collar region, the thermal emissions of the cloud tops show an evident decreasing trend from the evening towards morning hours visible at both southern and northern hemispheres with very good symmetry, especially noticeable at 3.8µm. This confirms both the average latitudinal symmetry reported by [Migliorini 2012] and [Haus 2013] and the local time variations that had only been studied separately for the northern [Taylor 1980] and southern cold collar [Grassi 2010, 2014]. Also in the cold collar region the opacity of the clouds starts to increase reducing the radiance seen through the atmospheric windows, consistent with the mean meridional circulation responsible of cloud material transport [Imamura and Hashimoto 1998] and [Sanchez-Lavega 2017], yielding to higher temperature and adiabatic compression of the vortex dynamics [Garate-Lopez 2015]. We also observe a major increase in the relative abundance of big particles at both hemispheres at high latitudes, confirming previous results by [Carlson 1993, Wilson 2008 and Haus 2013] with a potential increase in particle size from evening to morning hours, possibly explained by the dynamical forcing of the polar vortex and the downwelling inside the cold collar [Sanchez-Lavega 2017, Garate-Lopez 2018, Lebonnois 2018].

The detailed analysis of the polar areas is out of the scope of this study but our results confirm a higher cloud opacity, higher accumulation of big particle sizes and higher thermal brightness in the polar region up to at least 75º latitude in both hemispheres, consistent with the subsidence clouds depression reported in other works [Titov 2008, Ignatiev 2009, Cottini 2012, Cottini 2015] and the global atmospheric structure and dynamics summarized in the recent reviews by [Sanchez Lavega 2017], [Titov 2018] and [Limaye 2018b].

The detailed physical interpretation of these maps can be done unambiguously only with the support of a full general circulation model of Venus but at the same time this work can support in refining the main parameters used for the model itself, especially for persistent long term mean features.

## 7. ACKNOWLEDGEMENTS

We are grateful for the valuable comments provided by R. Hueso and another anonymous reviewer that have significantly improved the quality of this paper. We also thank our colleagues A. Longobardo, F. Tosi and G. Fillacchione for the useful discussions on limb darkening corrections and spectral radiance. This work has been funded by the European

Space Agency, the Italian Space Agency (ASI), French Space Agency (CNES) and other national agencies that supported the Venus Express mission and the VIRTIS instrument team.

## 8. DATA AVAILABILITY

All the infrared calibrated data from VIRTIS Venus Express used in this study is publicly available through the Planetary Science Archive hosted by the European Space Agency (https://archives.esac.esa.int/psa) in the datasets corresponding to the nominal mission phase and first mission extension: [VEX-V-VIRTIS-2-3-V3.0] and [VEX-V-VIRTIS-2-3-EXT1-V2.0]

ftp://psa.esac.esa.int/pub/mirror/VENUS-EXPRESS/VIRTIS/VEX-V-VIRTIS-2-3-EXT1-V2.0/